\documentclass[runningheads]{llncs}
\usepackage{graphicx}
\usepackage{hyperref}
\usepackage{xcolor}
\usepackage{amsmath}
\usepackage{subfig}
\usepackage{comment}
\usepackage{float}
\begin{document}
\title{ViDi: Descriptive Visual Data Clustering as Radiologist Assistant in COVID-19 Streamline Diagnostic}
\titlerunning{ViDi: Descriptive Visual Data Clustering for COVID-19 }
%
\author{Sahithya Ravi\inst{1}\orcidID{0000-1111-2222-3333} \and
Samaneh Khoshrou\inst{1}\orcidID{0000-0002-6317-9453} \and
Mykola Pechenizkiy\inst{1}\orcidID{0000-0003-4955-0743}}
%
%
\institute{Eindhoven University of Technology, Eindhoven, Netherlands\\
\email{\{s.ravi,s.khoshrou,m.pechenizkiy\}@tue.nl}}
\maketitle              
\begin{abstract}
In the light of the COVID-19 pandemic,  deep learning methods have been widely investigated in detecting COVID-19 from chest X-rays. However,  a more pragmatic approach to applying AI methods to a medical diagnosis is designing a framework that facilitates human-machine interaction and expert decision making. Studies have shown that categorization can play an essential rule in accelerating real-world decision making. Inspired by descriptive document clustering, we propose a domain-independent explanatory clustering framework to group contextually related instances and support radiologists' decision making. While most descriptive clustering approaches employ domain-specific characteristics to form meaningful clusters, we focus on model-level explanation as a more general-purpose element of every learning process to achieve cluster homogeneity. We employ DeepSHAP to generate homogeneous clusters in terms of disease severity and describe the clusters using favorable and unfavorable saliency maps, which visualize the class discriminating regions of an image. These human-interpretable maps complement radiologist knowledge to investigate the whole cluster at once. Besides, as part of this study, we evaluate a model based on VGG-19, which can identify COVID and pneumonia cases with a positive predictive value of 95\% and 97\%, respectively, comparable to the recent explainable approaches for COVID diagnosis.
\end{abstract}
\section{Introduction}
The COVID-19 outbreak is the biggest challenge of 2020, and even with all the efforts, the tolls are still rising around the world. Jam-packed hospital wards make efficient triage of patients with COVID-19 requisite \cite{cohen2020covid}. However, despite the rapid advancement of Artificial Intelligence (AI), AI in medicine is no silver bullet, and in most situations, domain experts still are sole decision-makers; COVID-related decisions are not an exception.
The top two reasons that make a complete AI-powered solution out of sights are first,\emph{``lack of representative and curated datasets''}, which may lead to train not only less effective but also unfair and biased models. Second, more data alone does not lead to a more practical model; as humans, we must understand and interpret the process as well as the outcome of an AI system \cite{vellido_2018}. \emph{``Lack of transparency and interpretability''} makes systems less trustworthy to be deployed in real-world situations \cite{10.5555/3364958}.
While a \emph{
complete AI-powered system entrusted with full responsibility acting autonomously on its own} is currently onerous, studies have shown that deployment of AI in human-in-the-loop fashion aiming to assist not substitute health-workers (especially radiologists) leading to better outcomes for patients \footnote{\url{https://medium.com/@zp489/f06e7daaee5}}. Today influx of COVID-19 patients and acute shortage of medical resources (i.e., staff and equipment) have highlighted the need for a collaborative human-AI diagnosis framework more than ever.


\paragraph{Main Contributions}In this work, we propose a visual data descriptive clustering method (ViDi) that aims to group CXR images with equal severity level and similar geographic extent of the infection together. The pipeline of ViDi starts with preprocessing and augmentation of more than 5000 CXR images, of which 200 are covid chest x-rays, followed by evaluation the success of the state-of-the-art VGGNet architecture \cite{vgg} in a transfer learning (TL) setting, on their proficiency in chest-xray detection. Then, class-discriminating saliency maps are generated using DeepSHAP\cite{shap}. ViDi helps the clinicians in two main ways: 1) Single image exploration: the framework provides two saliency maps. The \textit{favorable saliency map} highlights the regions that positively contribute to a particular prediction and the \textit{glum map} represents the areas that contribute negatively.  
 b) Cluster exploration: the framework groups CXR images by explanation similarity. Specifically, for COVID-19 dataset images with similar predicted severity scores and the infection's geographic extent are placed together. The clustering allows the clinician to compare CXR images based on different qualitative and quantitative measures and could be used to prioritize and adjust the treatment process, especially in overwhelmed circumstances. Besides, it provides a scalable solution for large scale CXR annotation. While existing works focus on predicting the severity scores from multiple data sources, including CXR images, here we focus on a framework that complements the findings from classic AI-based approaches and potentially facilitates and improve the decision-making process by radiologists or expert annotators.
\section{Background}\label{sec:background}
With a few exceptions \cite{DBLP:journals/corr/abs-1901-11210,DBLP:journals/corr/abs-2003-10769,DBLP:journals/corr/abs-2004-04582,DBLP:journals/corr/abs-2003-09871}, most AI-powered CXR analysis literature has focused on the accuracy of the prediction \cite{DBLP:journals/corr/abs-2003-10849,9069255,DBLP:journals/corr/abs-2003-11988,DBLP:journals/corr/abs-2003-12338,9069255} rather than enlightening the roots of the prediction. 
Ghoshal et al. \cite{DBLP:journals/corr/abs-2003-10769}  propose a Bayesian Convolutional Neural network to estimate the diagnosis uncertainty, which potentially yields more reliable prediction and can alert radiologists on false prediction.
DeepCovidExplainer \cite{DBLP:journals/corr/abs-2004-04582} highlights the class discriminative regions using a gradient-guided class activation maps; a heatmap is depicted to provide a human-interpretable explanation of the prediction. Since successive pooling layers have significantly decreased the deep activation maps' resolution, the class activation maps do not output a very precise localization.
Cohen et al. \cite{DBLP:journals/corr/abs-1901-11210} develop Chester, a web-delivered locally computed disease prediction system that aims to help clinicians understand a deep learning prediction. They use the gradient saliency map to explain network prediction. Practically speaking, the most discriminative (i.e., generally the gradient is high.) regions are depicted in red, while transparent regions have a negligible impact on the prediction. One issue with gradient-based interpreting approaches is that gradient is high at predictive regions and at locations that condition another region to impact, resulting in misguidance.
In another study, Wang \cite{DBLP:journals/corr/abs-2003-09871} proposed COVID-Net to detect distinctive abnormality in CXR images. The net employs GSInquire to identify the most critical regions (highlighted in red);  poor decision visualization is one of the main limitations of this approach. 
All the mentioned explainable frameworks share the same characteristics: they try to shed some light on network prediction in a single x-ray image to help clinicians in the decision making process. In contrast to most gradient-based methods, using a difference-from-reference, ViDi uses DeepLIFT, which allows propagating a quality signal even in situations where the gradient is zero and avoids artifacts caused by discontinuities in the gradient. In addition to highlighting the critical favorable and glum regions for a single image, our framework provides a cluster of similar images, which we believe has a great potential in streamline diagnosis of COVID-19.  Note that these methods complement radiologists' knowledge, and in fact, the domain expert is still the primary decision-maker.
\section{Experimental Design}\label{sec:experimentaldesign}
\subsection{Datasets \& Scenarios}\label{subsec:dataset}
We conduct our experiments on two public datasets: a) The COVID-19 x-ray image \cite{cohen2020covid}, including 392 COVID-19 postero-anterior chest x-ray images annotated with severity level. More information on the severity score assignment is available at \cite{wong2020covidnets}. b) the chest x-ray database from Kaggle \cite{kaggle}, consisting of 5863  chest x-ray images chosen from retrospective cohorts of pediatric patients graded by two expert physicians. 
We define three scenarios to evaluate the effectiveness of proposed framework:
\begin{enumerate}
    \item Binary classification of \textit{pneumonia vs normal} images of the pneumonia dataset.
    \item Binary classification of \textit{covid vs normal} images of the covid dataset.
    \item Multi class classification of \textit{mild vs medium vs severe} images of the covid dataset.
\end{enumerate}
The third task aims to classify the severity of the lung involvement based on the total opacity scores, ranging from 0 to 6. Similar to the grouping of lung involvement scores in COVID patients published by the Radiological Society of North America (RSNA) \cite{rsna}, we derive three categories mild, severe, and medium that refers to severity score below two, above four and within this range, respectively.
\subsection{Model, Pre-training, and Training}\label{sec:pre-training}
Since the COVID-19 image dataset is still small and evolving,  we employ a transfer learning strategy to utilize the knowledge from previously learned models and apply them to our problem. We trained VGG and DenseNet Architectures, followed by a prediction performance check. In this study, we choose VGG-19 model due to its success in the ImageNet Large Scale Visual Recognition Challenge \cite{imagenet} and COVID detection from chest x-rays as well as lighter computational complexity \cite{vgg}. For the pre-training phase, involving pneumonia vs. normal classification (scenario 1), we initialized the VGG-19 architecture with ImageNet weights and modified the classification layer.  By training the model on the pneumonia dataset, it learns common representations about lungs, which would be hard to achieve on the small set of COVID-19 images. The images used in both datasets were resized to 224 by 224 pixels and normalized using ImageNet standards. 

For the training phase, the pre-processed images of the COVID-19 dataset were randomly split into a train and test set in the ratio 70/30 for scenario 2 (covid vs normal) and 60/40 for scenario 3(severity classifier) such that the classes are stratified. In order to enhance the generalization capacity of the model, we employ image augmentation strategies such as random horizontal flip with a 50\% probability and randomized image rotation. The proposed model pretrained on the pneumonia dataset is then trained on the COVID19 dataset for both scenarios 2 and 3. The Adam optimizer a learning rate of 5e-4 was used for optimizing the weights. The experiments run for 50 epochs and batch size was set to 32. 
\vspace{-10pt}
\begin{figure}[!htbp]
  \centering
   \includegraphics[scale=0.4]{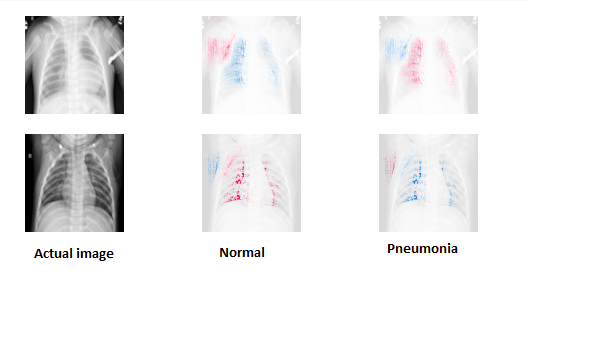}
   \vspace{-15pt}
  \caption{Glum and favorable saliency map using DeepSHAP}
  \label{fig:explain}
\end{figure}
\vspace{-10pt}
\section{Descriptive Visual Data Clustering} \label{sec:clustering}
 ``Descriptive clustering consists of automatically organizing data instances into clusters and generating a descriptive summary for each cluster. \cite{brockmeier_mu_ananiadou_goulermas_2018}''. The description should inform a user about each cluster's contents to speed up the decision-making process. There is no universal definition for a good description, and the selection of descriptions often relies on heuristic rules concerning the application.  We model descriptive clustering as categorizing similar CXR images with regard to the most critical regions (pixels) in network prediction.  The subset of features used to predict a class serves as its description in two ways: 1) Favorable regions that push the model towards a positive prediction. 2) Unfavorable (glum) pixels which contribute negatively to the prediction.
 
Once our model is trained, we employ DeepSHAP \cite{shap} to understand and visualize model predictions using saliency maps. We chose SHAP explanation, since SHAP values prove more consistent with human intuition than other methods \cite{shap} and they output a model agnostic explanation. DeepSHAP is an enhanced version of the DeepLIFT algorithm \cite{deeplift}, which measures feature contributions by calculating how a target($t$) changes as the input($x$) changes from the baseline. For a given input neuron $x$ with difference-from-baseline $\Delta x$ , and target neuron t with difference-from-baseline $\Delta t$ that we wish to compute the contribution to, DeepLIFT calculates the slope or multiplier given by, $m_{\Delta x \Delta t} = \frac {C{\Delta x \Delta t}} {\Delta x}$, where the DeepLIFT contribution scores are given by, $C{\Delta x \Delta t} = \Delta t$. DeepSHAP defines multipliers in terms of shapely values and works by recursively passing DeepLIFT multipliers via backpropagation. 
Figure \ref{fig:explain} illustrates an example saliency maps for two patients. For the first image of a pneumonia patient, blue regions in the lung with high opacity contribute negatively to the normal class. However, the same pixels contribute positively to pneumonia prediction (highlighted in red). While in the second image relating to a healthy patient, the red or positive shapely values correspond to transparent regions that contribute to the normal class.

Since shapely values try to isolate individual feature's effect, they are a good indicator of the similarity between instances. Clustering of images in shapely space can result in homogeneous clusters, with each cluster is dominated by instances with similar features. This is the main rationale behind our clustering algorithm. We pick K-means++ clustering due to its careful seeding method, simplicity, and speed in practice. The generated clusters are then visually inspected and assessed using clustering criteria in Sec. \ref{sec:results}.
\section{Results \& Discussion}\label{sec:results}
\subsection{Model performance}
The performance of the VGG-19 for three different scenarios is shown in Table \ref{tab:metrics}. The achieved recall and F1-score slightly outperforms \cite{DBLP:journals/corr/abs-2004-04582} with recall of 0.935 and F1-score of 0.928, and \cite{vgg} with a recall of 0.93. Out 65 COVID-19 patient samples, only 3 were misclassified as normal, which is analogous to the  misclassifcation ratio obtained in \cite{DBLP:journals/corr/abs-2004-04582}.

\begin{table}
\centering
\caption{Accuracy of the CNN on different settings}
\label{tab:metrics}
\resizebox{\textwidth}{!}{%
\begin{tabular}{|c|c|c|c|c|c|}
\hline
\textbf{Dataset} & \textbf{Model} & \textbf{Accuracy} & \textbf{F1 score} & \textbf{Precision} & \textbf{Recall} \\ \hline
pneumonia 2-class      & VGG-19 pretrained on ImageNet  & 0.94 & 0.89 & 0.83 & 0.97 \\ \hline
covid 2-class          & VGG-19 pretrained on pneumonia & 0.98 & 0.97 &  1    & 0.95   \\ \hline
covid severity 3-class & VGG-19 pretrained on pneumonia & 0.91 & 0.9 & 0.91 & 0.89   \\ \hline
\end{tabular}%
}
\end{table}

   \begin{figure}[!ht]
     \subfloat[Pneumonia vs normal\label{subfig-1:dummy1}]{%
       \includegraphics[width=0.3\textwidth]{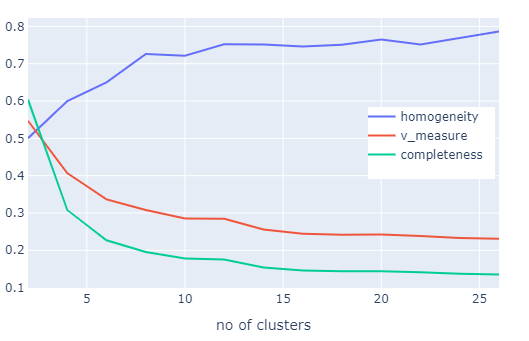}
     }
     \hfill
     \subfloat[Covid vs normal\label{subfig-2:dummy2}]{%
       \includegraphics[width=0.3\textwidth]{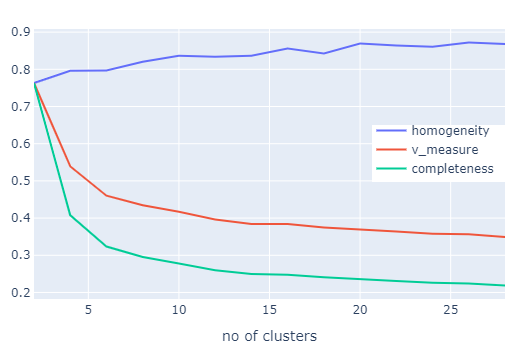}
     }
          \hfill
     \subfloat[Covid severity\label{subfig-2:dummy3}]{%
       \includegraphics[width=0.3\textwidth]{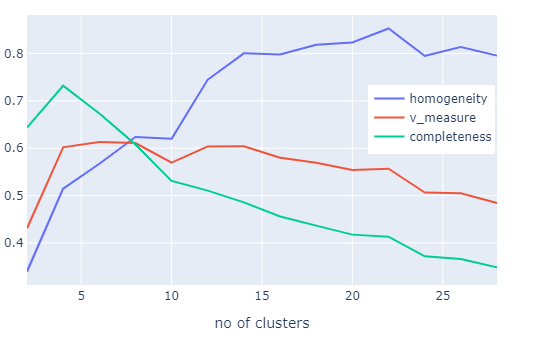}
     }
     \caption{Cluster quality assessment as a function of number of clusters (k)}
     \label{fig:dummy}
   \end{figure}
   \vspace{-10pt}
\begin{figure}[!htbp]
\begin{minipage}[b][][b]{.5\textwidth}
  \centering
  \includegraphics[scale=0.27]{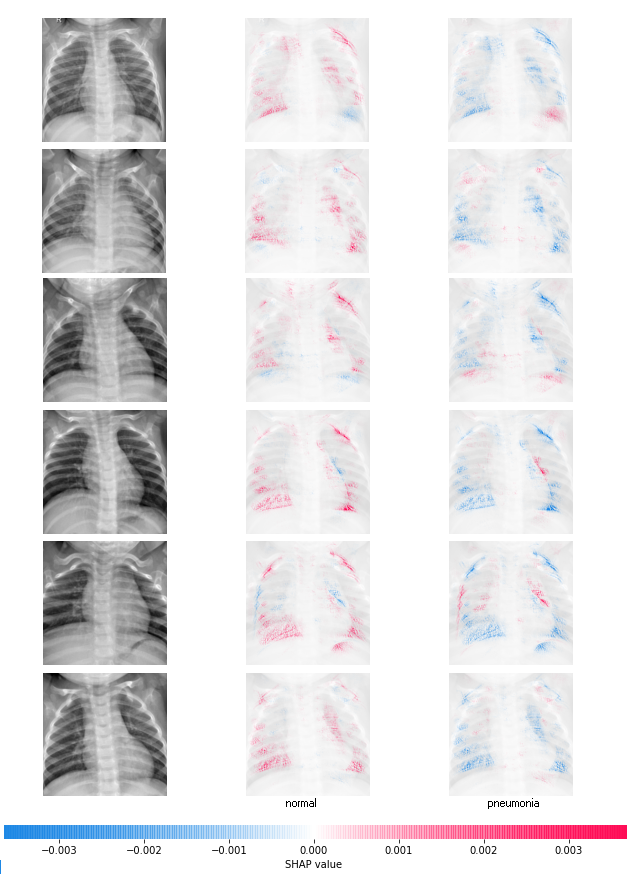}
  \caption{Normal pneumonia cluster}
  \label{fig:ex3}
  \end{minipage}%
\begin{minipage}[b][][b]{.5\textwidth}
  \centering
  \includegraphics[scale=0.3]{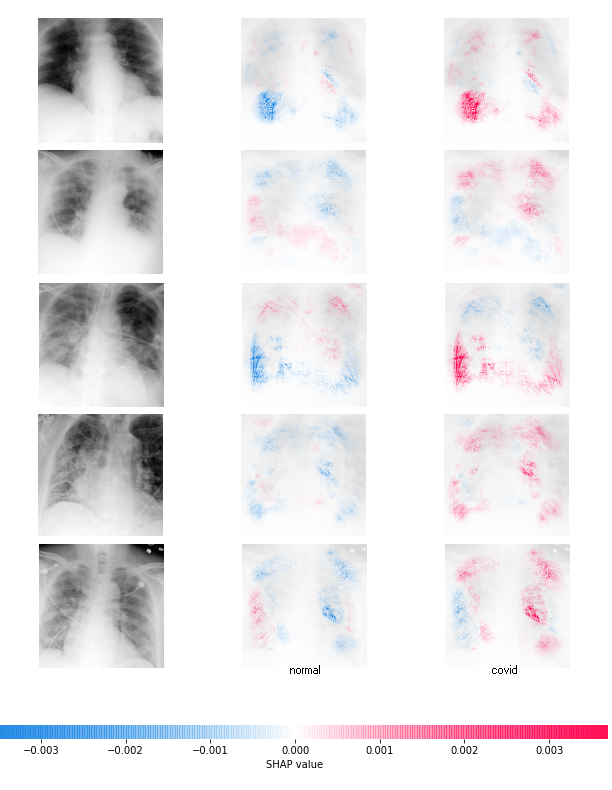}
  \caption{Covid cluster}
  \label{fig:covid1}
  \end{minipage}%
\end{figure}
\subsection{Clustering Analysis}
We evaluate the clustering algorithm based on homogeneity, completeness, and v-measure.
A clustering satisfies homogeneity if all of its clusters contain instances of a single class exclusively, whereas a clustering satisfies completeness when all instances of a given class are elements of a single cluster and not divided among many clusters. The v-measure is the harmonic mean of homogeneity and completeness.
The trade-off between homogeneity and completeness can be observed in all three scenarios, as shown in  Fig ~\ref{fig:dummy}.  While choosing the number of clusters, we prioritize homogeneity over completeness, as the number of clusters and the number of images per cluster at this stage are not too large to be handled by a domain expert. We choose k as 25, 20, and 22 for the three scenarios above, respectively.
\subsection{Cluster Visualization}
\begin{figure}[!htbp]
\begin{minipage}[b][][b]{.5\textwidth}
  \centering
  \includegraphics[scale=0.32]{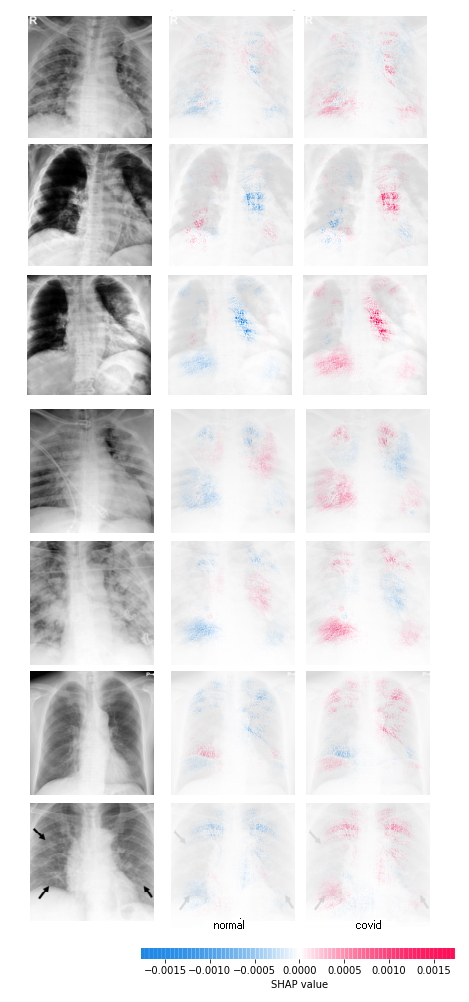}
  \caption{Impure covid cluster}
  \label{fig:impure}
  \end{minipage}%
\begin{minipage}[b][][b]{.5\textwidth}
  \centering
  \includegraphics[scale=0.38]{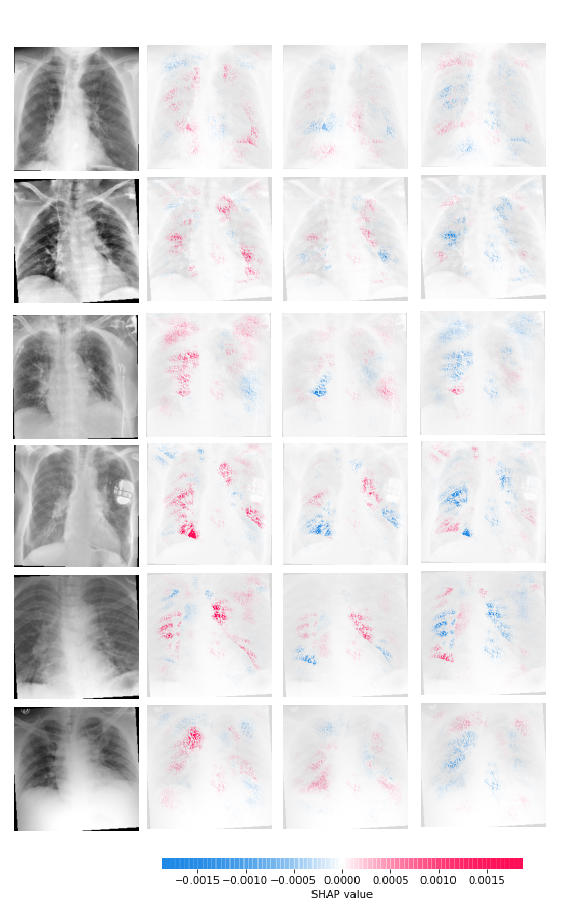}
  \caption{Mild covid cluster}
  \label{fig:mild}
  \end{minipage}%
\end{figure}

\begin{figure}[!htbp]
\begin{minipage}[b][][b]{.5\textwidth}
  \centering
  \includegraphics[scale=0.42]{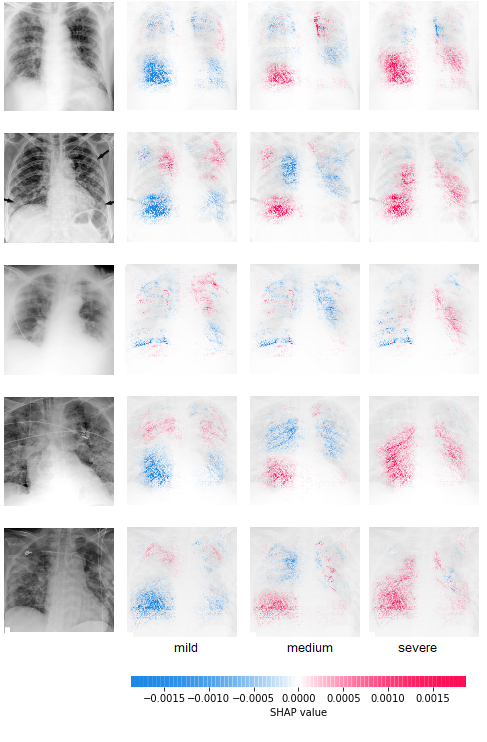}
  \caption{Severe covid cluster}
  \label{fig:sev}
  \end{minipage}%
\begin{minipage}[b][][b]{.5\textwidth}
  \centering
  \includegraphics[scale=0.42]{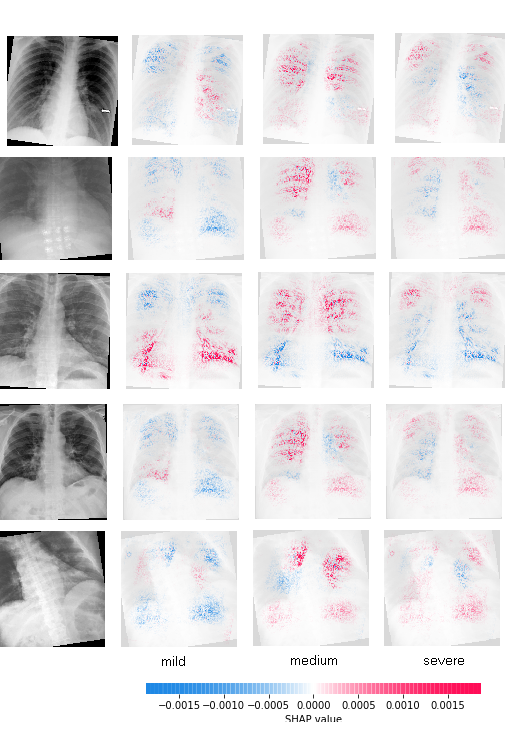}
  \caption{Moderate covid cluster}
  \label{fig:mod}
  \end{minipage}%
  \vspace{-5pt}
\end{figure}

In this section, we visualize examples of homogeneous clusters generated using the descriptive clustering algorithm for all three scenarios: 
\begin{enumerate}
    \item {\emph{Pneumonia vs Normal}} Figure~\ref{fig:ex3} exemplifies a homogeneous normal cluster and the corresponding saliency maps. The favorable (red) regions on the lungs denote the transparent regions contributing positively to normal class. For the pneumonia class, the same regions contribute negatively (blue).
    \item {\emph{Covid vs Normal}}: Figure~\ref{fig:covid1} illustrates a sample COVID-infected cluster. The favorable saliency map for COVID class and glum map for the normal class complement each other. For instance, in the first image, the opaque regions at the bottom of the lung contribute negatively to normal class and positively to the COVID class. Figure~\ref{fig:impure} shows an impure cluster generated from the COVID vs. normal scenario. All the cluster images are COVID positive, except for the fourth image from the top, which is a normal image. To achieve a higher number of pure clusters, it is thus important to determine the appropriate number of clusters (k).
    \item {\emph{Covid severity}}
Figure~\ref{fig:mild}, ~\ref{fig:mod} , and ~\ref{fig:sev} demonstrate a homogeneous mild, moderate and severe cluster respectively. Our approach provides two channels of information, why the model makes this prediction, and which regions push the model towards the other classes, which we believe would be more helpful than just providing a positive contribution. For example, in the severe cluster Figure ~\ref{fig:sev} , we have three maps highlighting the feature contribution in every three possible classes. It is a  cluster with maximum favorable (red) regions for severe class and prominent negative (blue) regions for mild class.  Comparing these maps facilitates false positives identification. Also, looking at the whole cluster potentially accelerates decision making.
\end{enumerate}
\section{Conclusion and Future Direction}
In this work, we present ViDi, a novel approach for model-based descriptive clustering with the help of explanation similarity. Rather than trying to replace a radiologist, ViDi has the potential to act as a bridge between AI and the expert, thus giving AI in a more convenient form to the medical community. In order to facilitate a better understanding of the generated clusters, we highlight both favorable and glum regions, which provide further insight into the behavior of the neural network. Further, the categorization of clusters into mild, moderate, and severe can act as a guiding mechanism for deciding the treatment option - at home, hospital, or ICU. We achieve high homogeneity of up to 80\% with the current version of the COVID-19 dataset, but deployment in a medical setting requires further enhancement of the framework. Since the current version of the dataset is small, a more in-depth analysis requires further data, especially in the severity front. As future work, cdvc zdv
\bibliographystyle{splncs04}
\bibliography{example_paper}
\end{document}